\documentclass[11pt,letterpaper]{article}  
\usepackage{fleqn,cospar}
\usepackage{url}
\usepackage{graphicx}

\usepackage[figuresright]{rotating}

\onecolumn

\title{ \vspace{11mm}
\bf CEM2K AND LAQGSM 
AS EVENT GENERATORS FOR SPACE-RADIATION-SHIELDING 
AND COSMIC-RAY-PROPAGATION APPLICATIONS}

\author{S.G.~Mashnik\address{Los Alamos National Laboratory, Los Alamos,  
NM 87545, USA},
K.K.~Gudima\address{Institute of Applied Physics, Academy of Science
of Moldova, Kishinev, MD-2028, Moldova}, 
I.V.~Moskalenko\address{NASA/Goddard Space Flight Center, Code 661, %
Greenbelt, MD 20771, USA}$^,$\address{Joint Center for Astrophysics,
University of Maryland, Baltimore County, Baltimore, MD 21250, USA},
R.E.~Prael$^1$,
and
A.J.~Sierk$^1$
}

\begin{document}

\maketitle

\begin{abstract}
The CEM2k and LAQGSM codes have been recently developed at
Los Alamos National Laboratory to simulate nuclear reactions for a
number of applications. We have benchmarked our codes
against most available 
data measured
at incident particle energies 
from 10 MeV to 800 GeV and have compared our results with predictions
of other current models used by the nuclear community. Here, we present
a brief description of our codes and show 
some
illustrative results 
that testify
that CEM2k and LAQGSM can be used as reliable event generators for
space-radiation-shielding, cosmic-ray propagation, and other astrophysical
applications. Finally, we show 
an example of combining of 
our calculated cross sections 
with experimental data from our LANL T-16
compilation to produce evaluated files. 
Such evaluated files were successfully used in the model of particle 
propagation in the Galaxy GALPROP 
to better constrain the size of the 
cosmic-ray
halo.
 
\end{abstract}

\section*{INTRODUCTION}
The radiation environment in deep space or on the moon can be very
harsh (MacFarlane et al., 1991). Without the Earth's atmosphere and 
magnetic field to protect them, astronauts will be directly exposed 
to high-energy galactic cosmic rays (GCR), to intense fluxes of protons 
and other particles from solar flares, and to radiation from on-board 
power plants or nuclear propulsion systems.
Such radiation is also very dangerous to computers and electronics on
spacecraft, as it may cause enough single-event upsets to lead to
device failure.
The economic penalties of additional mass to provide shielding 
may be 
huge. 
To minimize this expense, 
we need to be able to reliably calculate the amount of shielding 
required.
Performing this task
requires 
nuclear data for reactions induced by different projectiles, on different 
targets, for a large range of incident energies.
Similarly, 
addressing different astrophysical
problems, such as investigation the origin and  propagation  of cosmic
rays (CR) and the nuclide abundances in the solar system and in CR,
again 
requires 
a large amount of different nuclear reaction data.
Experiments to 
obtain such 
data are costly, there 
is
a limited
number of facilities available to make such measurements, 
and some reactions of interest cannot yet be measured at current
accelerators. Therefore reliable models and codes are required to
provide the necessary
estimates.

During recent years
at the Los Alamos National Laboratory, we have developed
an improved version of the 
Cascade-Exciton Model (CEM), 
contained 
in the code CEM2k, to describe 
nucleon-induced reactions at incident energies up to 5 GeV 
(Mashnik and Sierk, 2001 and 2002)
and the Los Alamos version of the Quark-Gluon String Model,
realized
in the high-energy code LAQGSM (Gudima et al., 2001),
able to describe both particle- and nucleus-induced reactions
at energies up to 1000 GeV/nucleon.
Both codes have been 
tested against
most of the available data 
and 
compared with predictions
of
other modern codes
(Mashnik and Sierk, 2001 and 2002; Mashnik et al., 2002a-2002d;
Titarenko et al., 2002). 
Our comparisons have shown that both codes describe 
a large variety of spallation, fission, and fragmentation reactions
quite reliably
and have one of the best predictive powers compared 
with 
other available 
Monte-Carlo codes. 
In the present paper, we outline our models 
and 
show several typical results
that testify that both CEM2k and LAQGSM are reliable event-generators which 
can be used in many different applications. As an illustration of application 
of our evaluated cross sections in astrophysics, we show an estimate of the 
size of the CR halo
obtained using the CR propagation code GALPROP
(Strong and Moskalenko, 1998; Moskalenko et al., 2002).

\section*{CEM2K AND LAQGSM}
A detailed description of the initial version of the CEM may be found
in Gudima et al. (1983), therefore we outline here only its basic
assumptions.
The CEM assumes that reactions occur in three stages. The first
stage is the IntraNuclear Cascade (INC) 
in which primary particles can be re-scattered and produce secondary
particles several times prior to absorption by or escape from the nucleus.
The excited residual nucleus remaining after the 
cascade determines the particle-hole configuration that is
the starting point for the preequilibrium stage of the
reaction. The subsequent relaxation of the nuclear excitation is
treated in terms of an improved Modified Exciton Model (MEM) of preequilibrium 
decay followed by the equilibrium evaporative final stage of the reaction.
Generally, all three stages contribute to experimentally measured outcomes.

The improved cascade-exciton model in the code CEM2k differs from 
the 
older
CEM95 version by incorporating new 
approximations for the elementary cross sections used in the cascade,
using more precise values for nuclear masses and 
pairing energies, 
employing a
corrected systematics for the level-density
parameters, improving the approximation for the pion ``binding energy", 
$V_{\pi}$, adjusting the cross sections for pion absorption on quasi-deuteron 
pairs inside a nucleus, considering the effects 
of refractions and reflections from the nuclear potential
on cascade particles,
allowing for nuclear transparency of pions, including the Pauli principle 
in the preequilibrium calculation, 
and
improving the calculation of the fission widths.
Implementation of 
significant refinements 
and improvements in the algorithms of many subroutines 
led to a decrease of
the computing time by up to a
factor of 6 for heavy nuclei, which 
is very important when performing
simulations with transport codes.
Essentially, CEM2k has a longer cascade stage,
less preequilibrium emission, and a longer evaporation stage
with a higher excitation energy, as compared to its precursors
CEM97 and CEM95.
Besides the changes to CEM97 and CEM95 mentioned above, we also made a 
number of other improvements and refinements, such as:
(i)
imposing momentum-energy conservation for each simulated event
(the Monte Carlo algorithm previously used in CEM 
provides momentum-energy conservation only 
statistically, on the average, but not exactly for the cascade stage 
of each event),
(ii)
using real binding energies for nucleons at the cascade 
stage instead of the approximation of a constant
separation energy of 7 MeV used in previous versions of the CEM,
(iii)
using reduced masses of particles in the calculation of their
emission widths instead of using the approximation
of no recoil used previously, and
(iv)
a better approximation of the total reaction cross sections.
On the whole, this set of improvements led to a much better description
of particle spectra and yields of residual nuclei and a better 
agreement with available data for a variety of reactions.
Details, examples, and further references may be found in
Mashnik and Sierk (2001 and 2002) and in Titarenko et al. (2002). 

The Los Alamos version of the Quark-Gluon String Model
(LAQGSM) by Gudima et al., (2001) is a next-generation of 
the Quark-Gluon String Model (QGSM) by Amelin et al.
(1990, and references therein) and is intended to describe
both particle- and nucleus-induced reactions at energies up to
about 1 TeV/nucleon. 
The core of the QGSM is built on a time-dependent version of the
intranuclear cascade model developed at Dubna,
often referred in the literature simply
as the Dubna intranuclear Cascade Model (DCM) (see Toneev and Gudima, 1983
and references therein).
The DCM models interactions of fast cascade particles (``participants")
with nucleon spectators of both the target and projectile nuclei and
includes 
interactions of two participants (cascade particles)
as well.
It uses experimental cross sections (or those calculated by the Quark-Gluon 
String Model for energies above 4.5 GeV/nucleon) for these
elementary interactions to simulate angular and energy distributions
of cascade particles, also considering the Pauli exclusion
principle. When the cascade stage of a reaction is completed, QGSM uses the
coalescence model described in Toneev and Gudima (1983)
to ``create" high-energy d, t, $^3$He, and $^4$He by
final state interactions among emitted cascade nucleons, already outside 
of the colliding nuclei.
After calculating the coalescence stage of a reaction, QGSM
moves to the description of the last slow stages of the interaction,
namely to preequilibrium decay and evaporation, with a possible competition
of fission
using the standard version of the CEM by Gudima et al.,
(1983). But if the residual nuclei have atomic numbers 
with  $A \le 13$, QGSM uses the Fermi break-up model 
to calculate their further disintegration instead of using
the preequilibrium and evaporation models.
LAQGSM differs from QGSM by replacing the preequilibrium and
evaporation parts  of QGSM described according to the standard CEM (Gudima
et al., 1983) with the new physics from CEM2k
(Mashnik and Sierk, 2001 and 2002) and has a number of improvements 
and refinements in the cascade and Fermi break-up models (in the
current version of LAQGSM, we use the Fermi break-up model only for
 $A \le 12$). A detailed description of LAQGSM and further
references may be found in Gudima et al. (2001).

Originally, both CEM2k and LAQGSM were not able to describe fission reactions 
and production of light fragments heavier than $^4$He, as they had neither 
a high-energy-fission nor a fragmentation model. 
Recently, we addressed 
these problems 
(Mashnik et al., 2002a and  2002c) by further improving
our codes and by merging them with the Generalized Evaporation Model
code GEM2 developed by Furihata (2000 and 2001).

We have benchmarked our codes on all reactions measured recently at 
GSI (Darmstadt, Germany) 
and on many other different
reactions at lower and higher energies measured 
earlier at other laboratories. We found that CEM2k and LAQGSM allow us to 
describe quite well a large variety of spallation,
fission and fragmentation reactions at energies
from 10 MeV to 800 GeV, without any free 
fitting parameters,
though we still have to solve a number of problems for a better
description of nuclides produced near the intersection of the spallation 
and fission 
regions. 
Our current versions of CEM2k and LAQGSM were
incorporated recently into the MARS (Mokhov, 1995) and LAHET
(Prael and Lichtenstein, 1989) transport codes and are currently
being incorporated into MCNPX (Waters, 1999).
 This will allow others to use our codes as event-generators
in these transport codes to simulate reactions with targets of
practically arbitrary geometry and nuclide composition.

Figure 1 shows examples of calculated by CEM2k and LAQGSM 
neutron spectra from interactions of protons with $^{208}$Pb at 
0.8 GeV, while Figure 2 gives examples of p, d, t,
$^3$He, and $^4$He spectra from p (61 MeV) + Fe compared with experimental 
data by Ishibashi et al. (1997) and Bertrand and Peelle (1973). 

\begin{figure*}[!b]
\begin{minipage}[b]{88mm}
\hspace{-14mm}
\includegraphics[width=80mm,angle=-90]{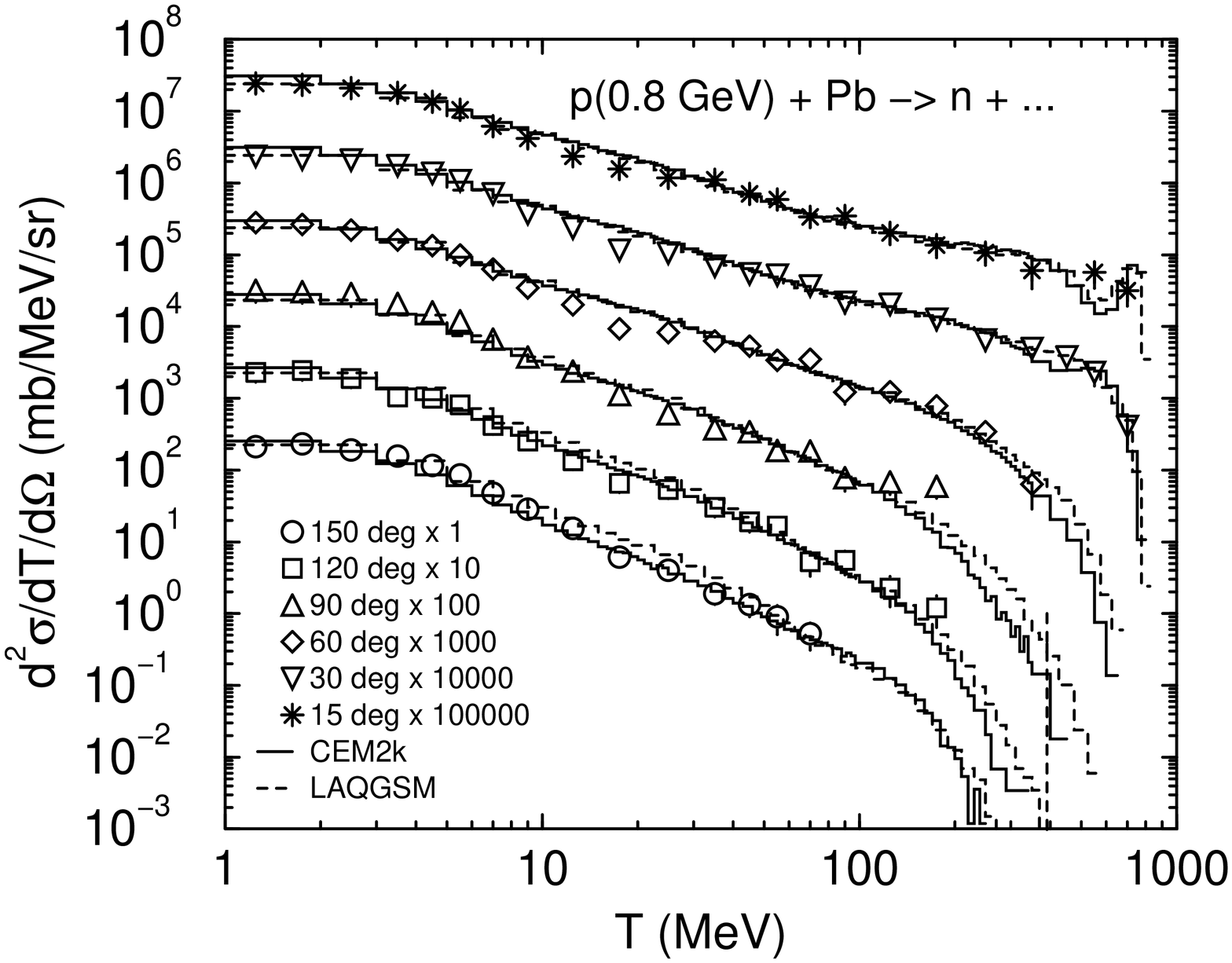}
\vspace{-11mm}
\caption{
Comparison of measured (Ishibashi et al., 1997) double differential cross
sections of neutrons from 0.8 GeV protons on Pb with
CEM2k and LAQGSM calculations. 
}
\end{minipage}
\hspace{\fill}
\begin{minipage}[b]{88mm}
\hspace{-14mm}
\includegraphics[width=80mm,angle=-90]{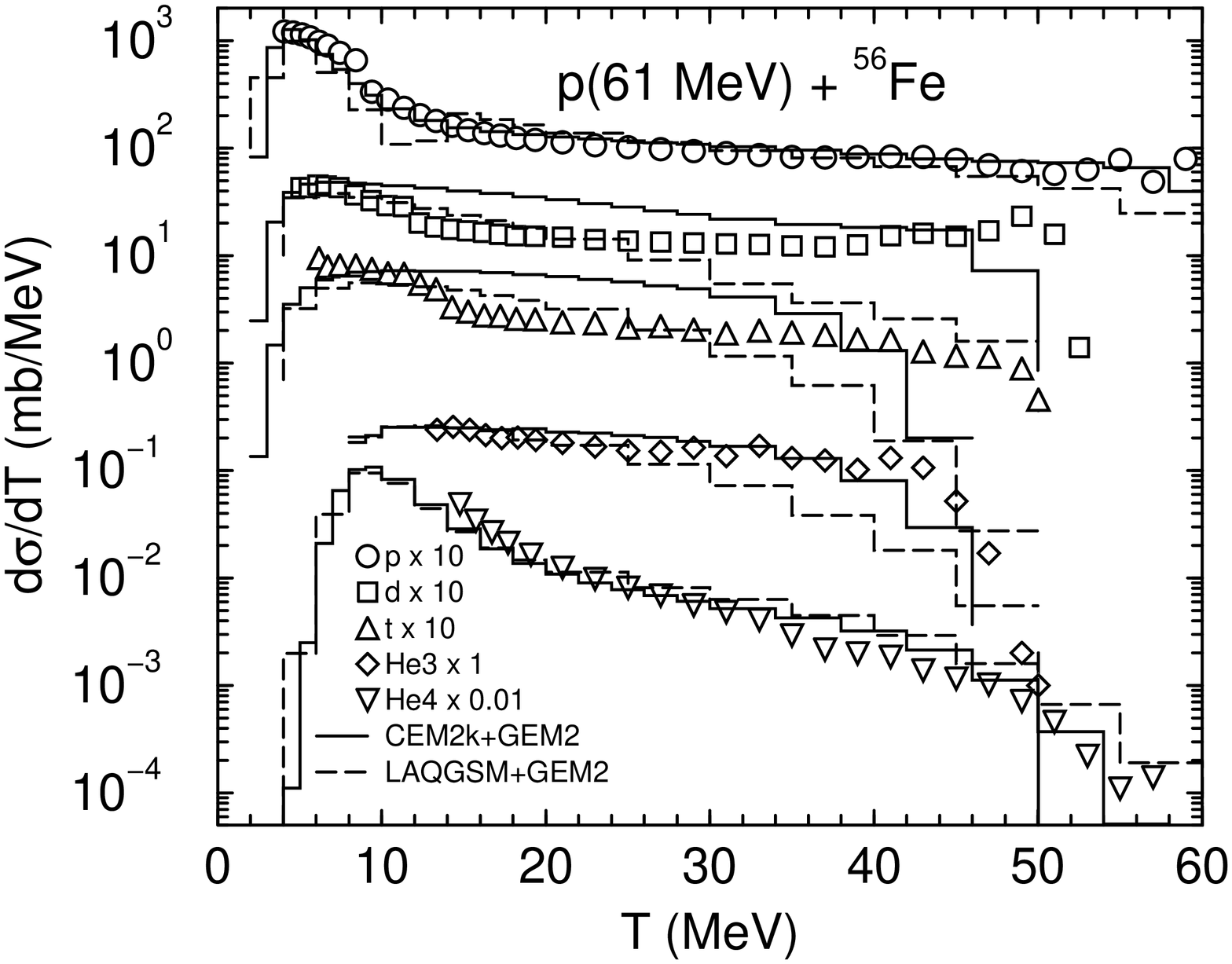}
\vspace{-11mm}
\caption{
Comparison of measured (Bertrand and Peelle, 1973) angle-itegrated
energy spectra of p, d, t, $^3$He, and $^4$He from 61 MeV
protons on Fe with
CEM2k+GEM2 and LAQGSM+GEM2 calculations. 
}
\end{minipage}
\end{figure*}

We note that all reactions shown in Figures 1 and 2, and in all the  following
figures of this paper, were calculated within a single approach, without 
fitting any parameters of CEM2k or LAQGSM. 
We see that both CEM2k and LAQGSM describe well spectra of secondary 
neutrons and protons. Similar results are obtained for other targets and 
energies of incident protons, as well as for reactions induced by neutrons,
pions, and photons.
Spectra of  $^4$He are also described 
by our codes
quite well,
while
spectra of d, t, and $^3$He are reproduced reasonably, but not 
as well as those of 
nucleons and $^4$He. This is partially due to the fact that we
do not fit here the probability $\gamma_j$ for $p_j$ excited nucleons
(excitons) to condense into a complex particle which can be emitted
during the preequilibrium stage of a reaction
to get the best agreement with the data, as is often done in the
literature by other authors (see details in Gudima et al., 1983 
and Mashnik et al., 2002c). In the present version of
our models we do not take into account direct production of complex particles
like pick-up and knock-out, and this leads to some underestimation
of emission of high-energy complex particles, and to under-prediction of
the high-energy tails of their spectra seen in Figure 2. We note that 
some of the evaporation models used often in the literature, like the
GSI evaporation model by Schmidt (Gaimard and Schmidt, 1991; Junghans 
et al., 1998), which is used in conjunction with the Liege INC by Cugnon 
(Boudard et al., 2002) in the code INCL,
do not consider preequilibrium particle emission;
they only evaporate n, p, and $^4$He, and do not produce
d, t, and $^3$He at all.

\begin{figure*}[!b]
\begin{minipage}[b]{88mm}
\vspace{-10mm}
\includegraphics[width=90mm]{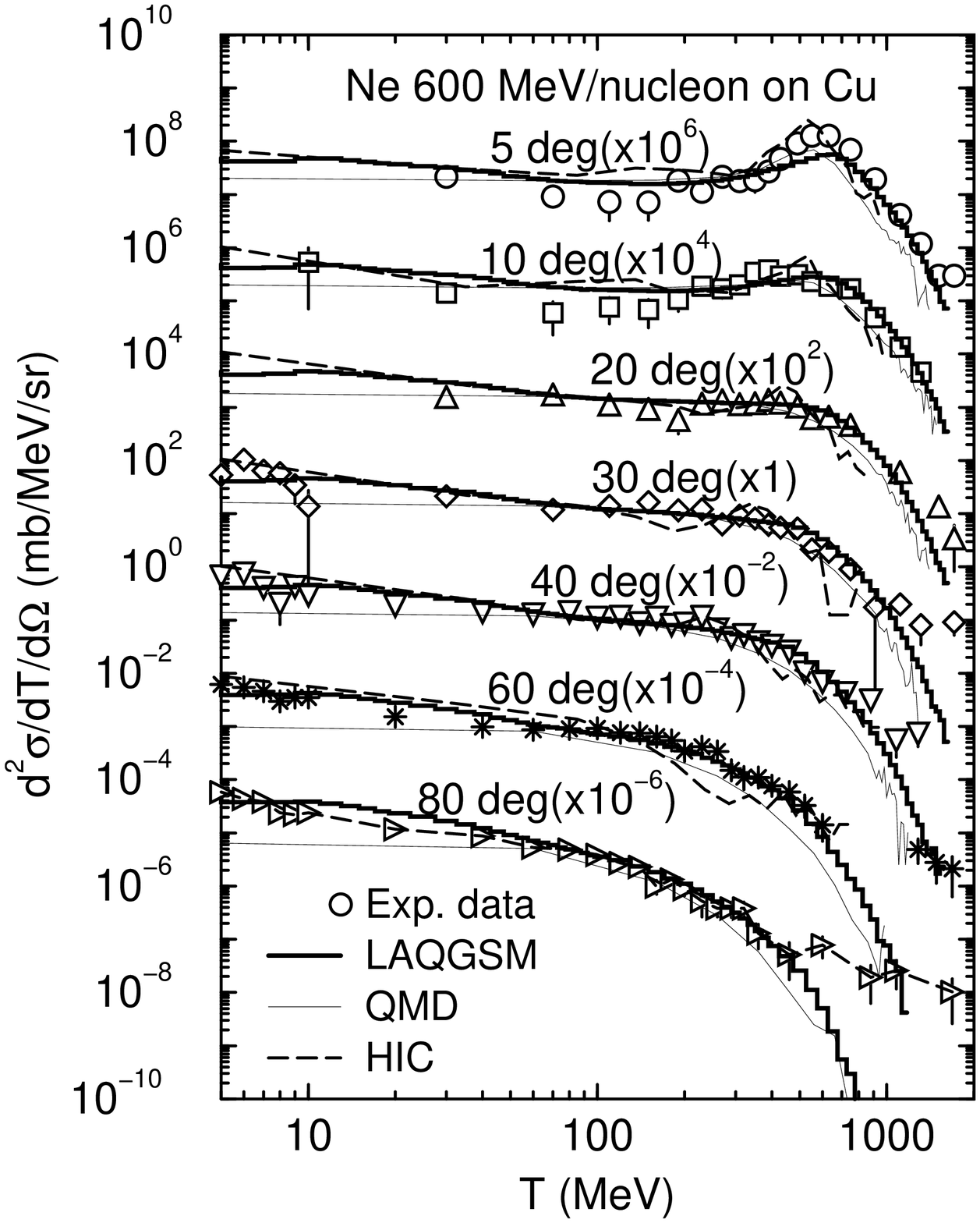} 
\vspace{-20mm}
\caption{
Comparison of measured (Iwata et al., 2001) double differential cross
sections of neutrons from 600 MeV/nucleon $^{20}$Ne on Cu with our
LAQGSM results and calculations by QMD and HIC (Iwata et al., 2001).
}
\end{minipage}
\hspace{\fill}
\begin{minipage}[b]{88mm}
\vspace{-10mm}
\includegraphics[width=95mm]{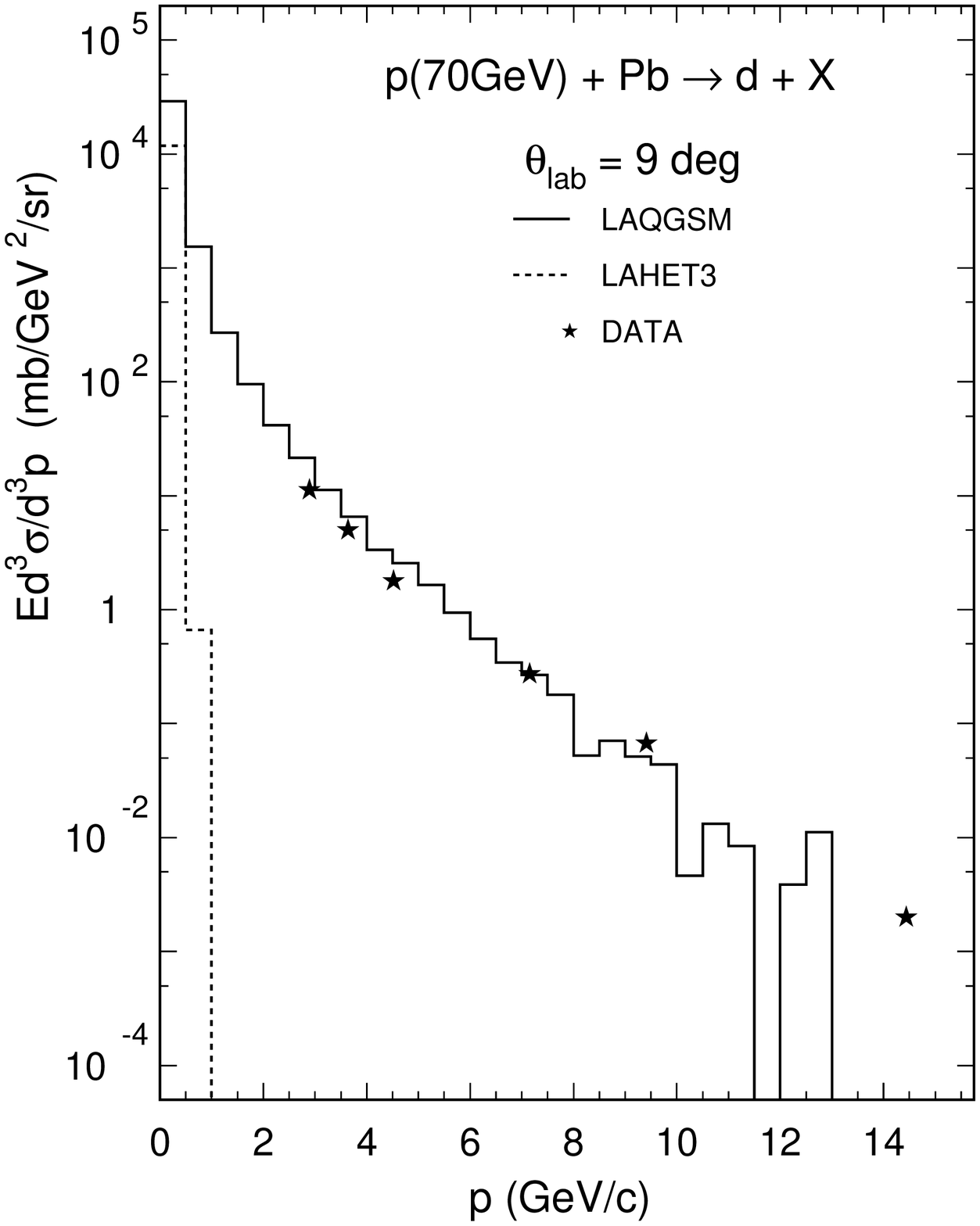} 
\vspace{-20mm}
\caption{
Invariant cross section for the production 
of d
by 70 GeV protons on Pb at 
9.17$^\circ$
as a function of deuteron momentum. Experimental data
(Abramov et al.,1987) and our calculations with LAQGSM and LAHET3
are shown as indicated in 
the
legend.
}
\end{minipage}
\end{figure*}

Recently, Nakamura's group measured neutron double-differential
cross sections from many reactions induced by light and medium nuclei
on targets from $^{12}$C to $^{208}$Pb, at several incident energies
from 95 to 600 MeV/nucleon (Iwata et al., 2001 and references therein).
We have calculated all these cross sections using LAQGSM.
As an example, 
our results for interactions of 600 MeV/nucleon $^{20}$Ne with
Cu are  compared in Figure 3 with experimental data and calculations
with the QMD (Aichelin, 1991) and HIC (Bertini et al., 1974)
models kindly provided to us by Nakamura's group.
We see that LAQGSM describes these data quite well and agrees with the 
measurements much better than do QMD and HIC. Similar results are
obtained for all the other reactions measured by this group.

Recently, 
for the Proton Radiography (PRAD) project,
we have performed (Mashnik et al., 2002b) 
a benchmark of QGSM (and LAQGSM), MARS,
and LAHET3 (Prael, 2001)  against all measured double-differential
cross sections 
of
proton-nucleus reactions at energies around 50 GeV 
for all
targets for which we were able to find experimental data.
We calculated, compared with experimental data,
and analyzed more than 500 spectra of p, d, t, $^3$He,
$^4$He, $\pi ^-$, $\pi ^+$, $K^-$, $K^+$, and $\bar p$ emitted from
targets from $^9$Be to $^{208}$Pb at angles from 0 to 159 deg at
incident proton energies from 30 to 70 GeV.
This study has shown that LAQGSM describes
well most of the measured spectra.
A detailed and
comprehensive report on this work is now in preparation,
while
Figure 4 shows just
one example, namely spectra of deuterons emitted at 9.17 deg from 
p (70 GeV) + $^{208}$Pb.
We see that deuterons with momentum  
of
up to about 
15 GeV/c are emitted and measured in this particular reaction. 
Utilizing the coalescence mechanism for complex particle emission, LAQGSM
is able to describe high-energy deuteron production, and agrees well with
the measurement. 
LAHET3
does not consider
the coalescence of complex particles and therefore describes emission of
only evaporative and preequilibrium deuterons with momenta not 
higher than 1 GeV/c. 
Even though the cross section
for emission of deuterons with momentum 
of
$\sim 15$ GeV/c
is more than six orders of magnitude lower than for evaporation of 
low energy deuterons, such high energy deuterons and other complex
particles may be extremely dangerous to people and equipment in space.
Therefore, we need to be able to calculate
such reactions as well as possible, to accurately estimate the
necessary shielding.

Recently at GSI in Darmstadt, Germany, a large amount of measurements 
have been performed using inverse kinematics
for interactions of $^{56}$Fe, $^{208}$Pb and $^{238}$U
at 1 GeV/nucleon and $^{197}$Au at 800 MeV/nucleon with 
liquid $^1$H. These measurements  provide
a very rich set of cross sections for production of practically
all possible isotopes from such reactions in a ``pure" form,
{\it i.e.}, individual cross sections from a specific given bombarding isotope
(or target isotope, when considering reactions in the usual kinematics,
p + A). Such cross sections are much easier to compare to models than the 
``camouflaged" data from $\gamma$-spectrometry measurements. These 
are often obtained only for a natural composition of isotopes in a target
and are mainly for cumulative production, where measured cross sections
contain contributions not only from the direct production of
a given isotope, but also from all its decay-chain precursors. 
In addition, many reactions where a beam of light, medium,
or heavy ions with energy near to or below 1 GeV/nucleon interact with 
different nuclei, from the lightest, d, to the heaviest, $^{208}$Pb
were measured recently at GSI. References on these measurements and
many tabulated experimental cross sections may be found on the Web
page of Prof. Schmidt (2002). We have analyzed with CEM2k and LAQGSM 
all measurements done at GSI we are acquainted of, both for
proton-nucleus and nucleus-nucleus interactions. Some examples of
our CEM2k results compared with the GSI
data and calculations by other current models
for proton-nucleus reactions may be found in
Mashnik and Sierk (2001 and 2002), Mashnik et al. (2002a, 2002c, and 2002d), 
and
Titarenko et al. (2002). Figure 5 shows an example of LAQGSM
results for the reaction p(1 GeV) + $^{208}$Pb compared with
GSI data and calculations by a version of 
LAHET using the INCL event-generator
incorporated recently into LAHET.  We have described INCL above;
it is well known and often used in Europe.
One can see that LAQGSM merged with GEM2 (LAQGSM+GEM2)
describes quite well the GSI data and agrees better with the measurement
than INCL does. Similar results are obtained for all other
reactions measured at GSI for which we found data.

\begin{figure*}[!tb]
\vspace*{-1cm}
\begin{minipage}[b]{88mm}
\hspace{-14mm}
\vspace{-2\baselineskip}
\includegraphics[width=83mm,angle=-90]{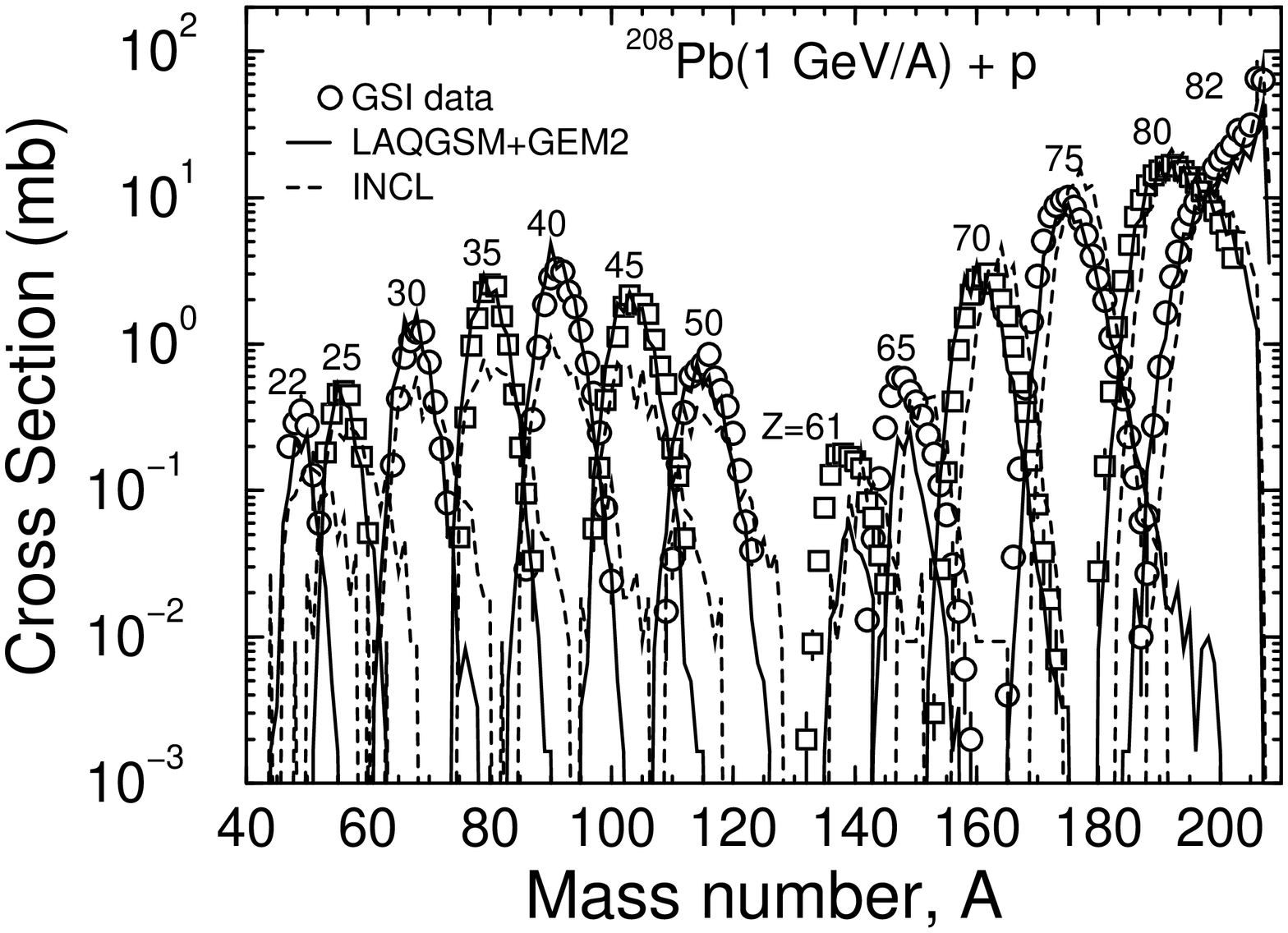} 
\caption{
Experimental (Enqvist et al., 2001) mass distributions of the cross sections 
of thirteen isotopes with the charge $Z$ from 22 to 82
produced in the reaction p(1 GeV) + $^{208}$Pb
compared with our LAQGSM+GEM2 calculation and the INCL code (see the text
for a description).
\vspace{0.8\baselineskip}
}
\end{minipage}
\hspace{\fill}
\begin{minipage}[b]{88mm}
\hspace{-8mm}
\includegraphics[width=103mm]{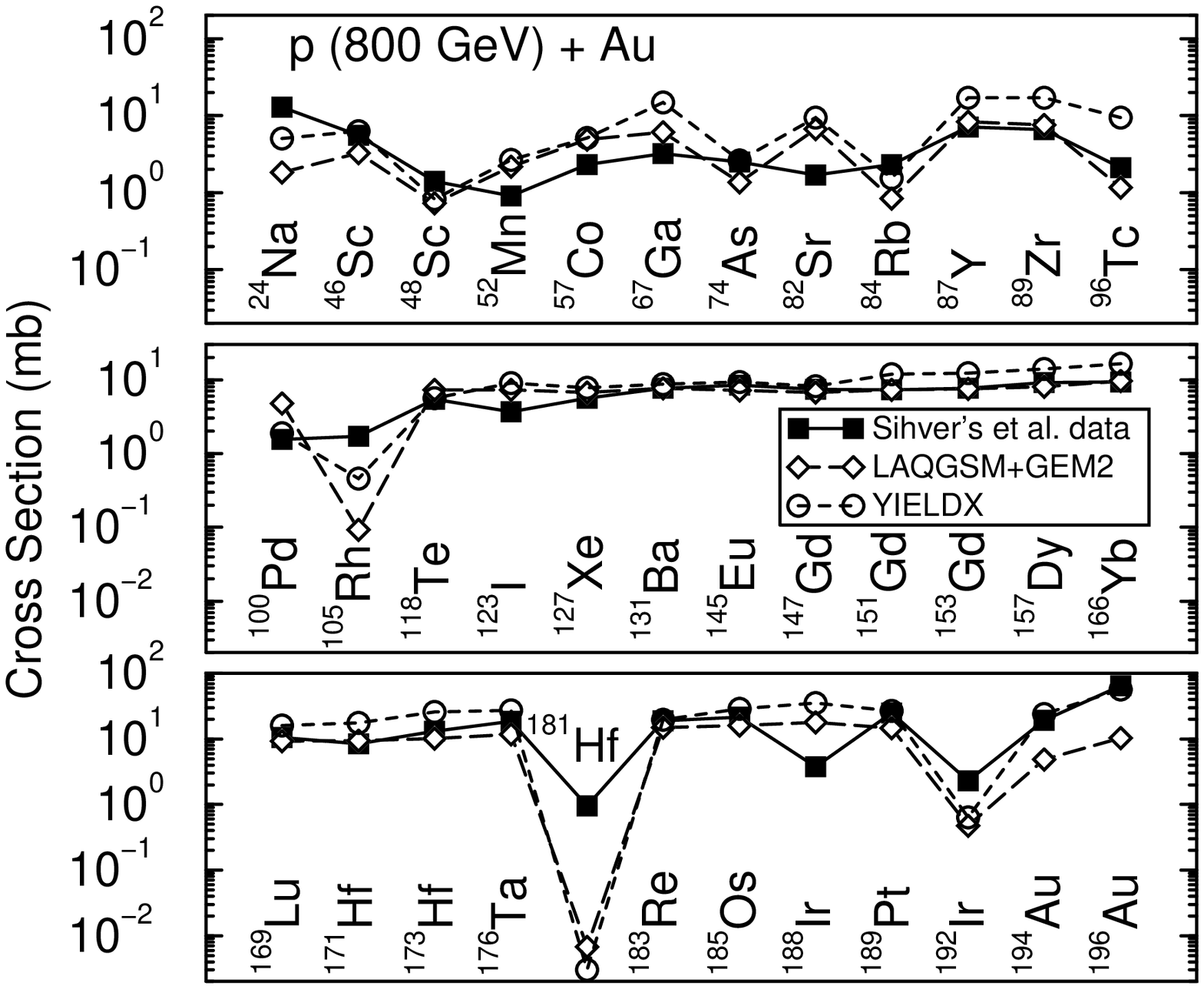} 
\vspace{-70mm}
\caption{
Detailed comparison between experimental (Sihver et al., 1992)
and calculated cross sections using LAQGSM+GEM2 and
YIELDX (Silberberg et al., 1998)
Only every third measured product from the reaction p(800 GeV) + Au
is plotted. All cross sections shown here
except for $^{46}$Sc, $^{74}$As, $^{96}$Tc, $^{188}$Ir,
and $^{192}$Ir are cumulative.
}
\end{minipage}
\end{figure*}

Figure 6 shows an example of 
the highest energy we calculated
so far with LAQGSM, namely
results by LAQGSM+GEM2 for the reaction  p(800 GeV) + A 
compared with
experimental data (Sihver et al., 1992)
and calculations with the phenomenological code
YIELDX (Silberberg et al., 1998). 
For the sake of brevity, we present in this figure only
every third yield measured and tabulated in Sihver et al. (1992),
though we calculated all possible products from this reaction and
get similar results for nuclides not shown here. One can see that
LAQGSM agrees reasonably with most of the measured yields and describes
the data better than YIELDX.
More than a half of the measured products are described by LAQGSM+GEM2
with an accuracy of a factor of two or better, though we find some
large 
discrepances for several nuclides like $^{181}$Hf and $^{105}$Rh.

At last, Figure 7 shows a heavy-ion induced reaction measured at GSI
(Junghans, 1997; Junghans et al., 1998), namely the yields
of measured products (black circles) from the interaction
of a 950 MeV/nucleon $^{238}$U beam with copper compared with our LAQGSM+GEM2
results (open circles). One can see that LAQGSM+GEM2 describes most of these
data with an accuracy of a factor of two or better. 

\section*{HALO SIZE LIMITS FROM THE GALPROP MODEL}
This section shows an example of using cross sections calculated by CEM2k 
and LAQGSM (together with available data) 
to put constraints on the Galactic CR halo size.
For this calculation we use a state-of-the-art 
CR propagation code 
GALPROP\footnote{GALPROP model including software and data sets is 
available at \url{http://www.gamma.mpe-garching.mpg.de/~aws/aws.html}}.
The GALPROP models have been described in full detail elsewhere
(Strong and Moskalenko, 1998; Moskalenko et al., 2002 and references therein); 
here we summarize their basic features.

The code 
solves a transport equation
on a full 3D spacial grid $(x,y,z)$ or 2D grid.
The 2D models have cylindrical symmetry in the
Galaxy, and the basic coordinates are $(R,z,p)$, where $R$ is
Galactocentric radius, $z$ is the distance from the Galactic plane and
$p$ is the total particle momentum. The  propagation
region is bounded by $R=R_h$ (taken $30$ kpc), 
$z=\pm z_h$ beyond which free escape is
assumed. For a given $z_h$ the diffusion
coefficient as a function of momentum  and the reacceleration
parameters is determined by the energy-dependence of the B/C ratio.  
The spatial diffusion coefficient is taken as
$\beta D_0(\rho/\rho_0)^\delta$, assuming independence of position, 
where $\rho$ is rigidity. For the case of
reacceleration the momentum-space diffusion coefficient $D_{pp}$ is
related to the spatial coefficient.
The reacceleration is parameterized by $v_A^2/w$
where $v_A$ is the Alfv\'en speed and $w$ the ratio of wave energy 
density to magnetic field energy density.
The 
source
spectrum of nuclei is assumed
to be a power law in momentum, $dq(p)/dp \propto p^{-\gamma}$ for the
injected particle density, if necessary with a break.

The interstellar hydrogen distribution uses H~{\sc i} and CO surveys and
information on the ionized component; the Helium fraction of the gas is
taken as 0.11 by number.  Energy losses of
nuclei by ionization and Coulomb interactions
are included. 
The distribution of CR
sources is chosen to reproduce the CR distribution
determined by analysis of EGRET gamma-ray data.
The primary source abundances are adjusted to give as good agreement 
as possible with the observed abundances after propagation, 
for a given set of cross-sections.
The heliospheric modulation is taken into account using the force-field 
approximation.

The nuclear reaction network
is built using the Nuclear Data Sheets. 
The isotopic cross section database now includes
the LANL T-16 compilation
by Mashnik et al. (1999) consisting of
several tens of thousands of experimental points.
This includes a critical re-evaluation of some data and cross checks.
The isotopic cross sections are calculated in GALPROP
using the authors' fits to major beryllium and boron 
production cross sections C,N,O $\to$ Be,B, and to other major reactions;
all
other cross sections 
at present
are calculated using 
the Webber et al.\ (1990)
and/or Silberberg et al. (1998)
phenomenological approximations renormalized to the data where it exists.
The reaction network is solved starting at the heaviest nuclei 
(i.e.\ $^{64}$Ni), solving the propagation equation, computing all 
the resulting secondary source
functions, and proceeding to the nuclei with $A-1$. 
The procedure is repeated down to $A=1$.
In this way all secondary, tertiary etc.\ reactions are 
automatically accounted for.
To be completely accurate for all isotopes, e.g.\ for some rare 
cases of $\beta^\pm$-decay, the whole loop is repeated twice.

\begin{figure*}[!tb]
\vspace{-40mm} 
\hspace{-5mm}
\centerline{\includegraphics[width=210mm]{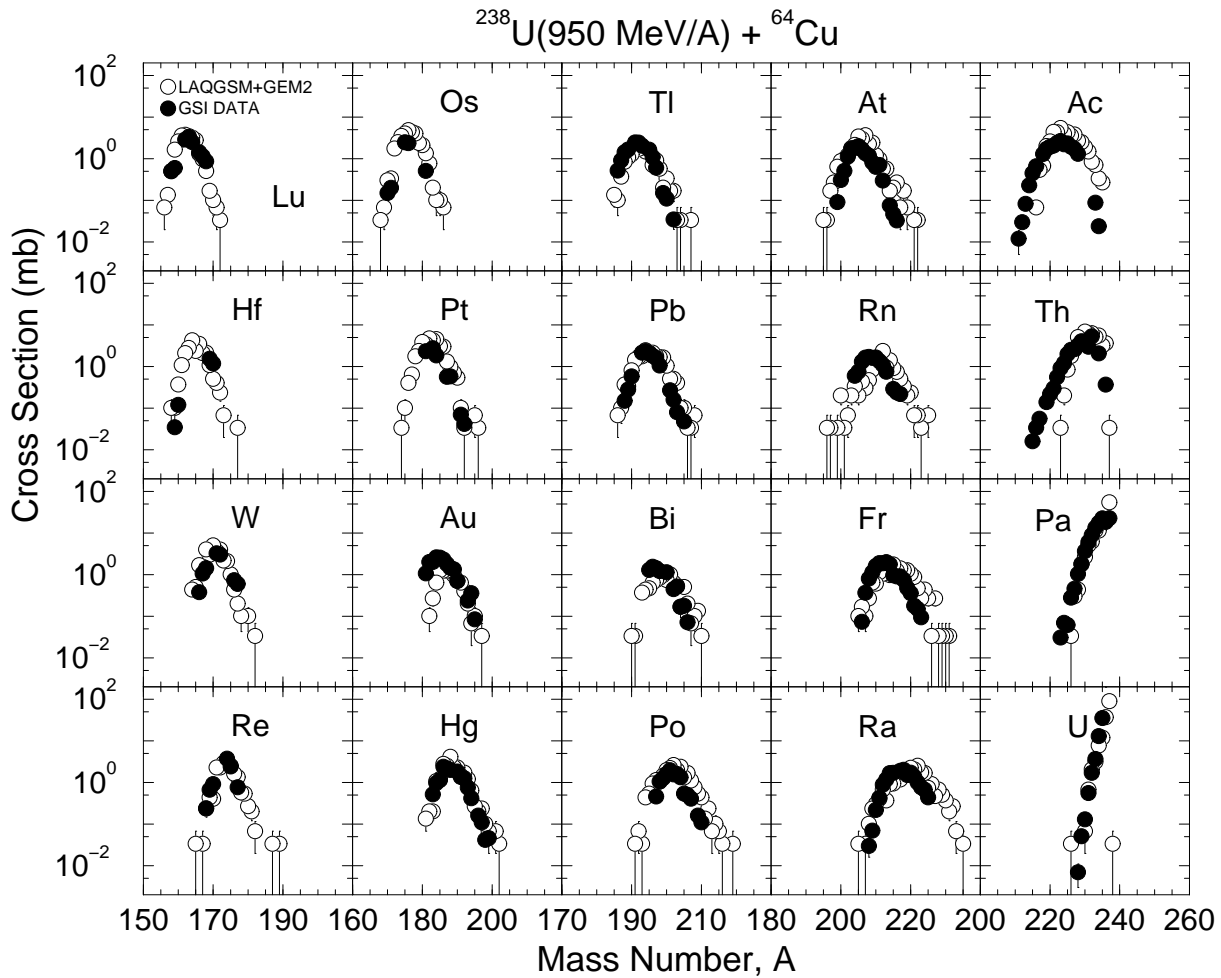}} 
\vspace{-140mm}
\caption{
Comparison of measured (Junghans, 1997; Junghans et al., 1998)
cross sections of projectile fragments produced
by a 950 MeV/nucleon $^{238}$U beam in a copper target (black circles)
 with our LAQGSM+GEM2 results (open circles).
}
\end{figure*}

For some astrophysically important reactions we produced
our own evaluations of excitation functions
(e.g., Moskalenko et al., 2001) instead of using
only scarce experimental data or calculations by stand-alone
phenomenological systematics or nuclear reaction models.
For this purpose we used all available to us experimental 
data from the LANL T-16 compilation (Mashnik et al., 1999) together with 
calculations by CEM2k, and for several reactions, by LAQGSM and the older 
versions of the CEM code, CEM97 and CEM95.
One example of such an evaluated excitation function, for
the reaction $^{nat}$Si(p,x)$^{26}$Al, is shown in Figure 8
together with available data, CEM2k results, 
and calculations using
phenomenological systematics by Webber et al.\ (1990) and
Silberberg et al.\ (1998). It is seen that CEM2k
has some problems in a correct description of this
particular cross section near the threshold; 
therefore we used abundant experimental
data available for this reaction to produce our evaluated
excitation function at these energies. It is clear 
that neither the Webber et al.\ (1990) systematics nor the Silberberg
et al.\ (1998) approximation describe correctly this excitation function;
using their results as an input to CR propagation codes 
may lead to errors in results and interpretation.

The results of the calculation of Galactic propagation of
radioactive isotopes $^{26}$Al, $^{36}$Cl, and $^{54}$Mn are shown in Figure 9,
where
the evaluated excitation functions used
were produced as described above.
The radioactive isotopes of these elements
are the main astrophysical ``time clocks'' which together with stable
secondary isotopes allow us to probe global Galactic properties, 
such as the diffusion coefficient and the halo size.
Based on the CR data from spacecraft (ACE, Ulysses, and Voyager,
for details see Moskalenko et al., 2001) we were able to 
restrict the halo size as $z_h\sim4-6$ kpc.
Using the semiempirical 
systematics 
yields less consistent
results (see, e.g., Strong and Moskalenko, 2001) rizing 
questions about the interpretation. This result supports
the conclusion that large uncertainties 
for the halo size obtained in previous works were mostly
due to cross-section inaccuracies. 

In future, we plan to develop evaluated data libraries for 
other astrophysical reactions of interest
and to use them in future studies of Galactic CR propagation.

\section*{SUMMARY}
From the results presented here and in the cited references,
we conclude that CEM2k and LAQGSM describe well
(and
without any
fitting parameters) a large variety of medium- and high-energy nuclear
reactions and are suitable for evaluations of nuclear data for
science and applications. We continue our work on further improvements
and development of both
CEM2k and LAQGSM, but even in their present versions they are
quite reliable and may be used as event-generators for astrophysical
applications.

\begin{figure*}[!t]
\begin{minipage}[t]{88mm}
\includegraphics[width=88mm]{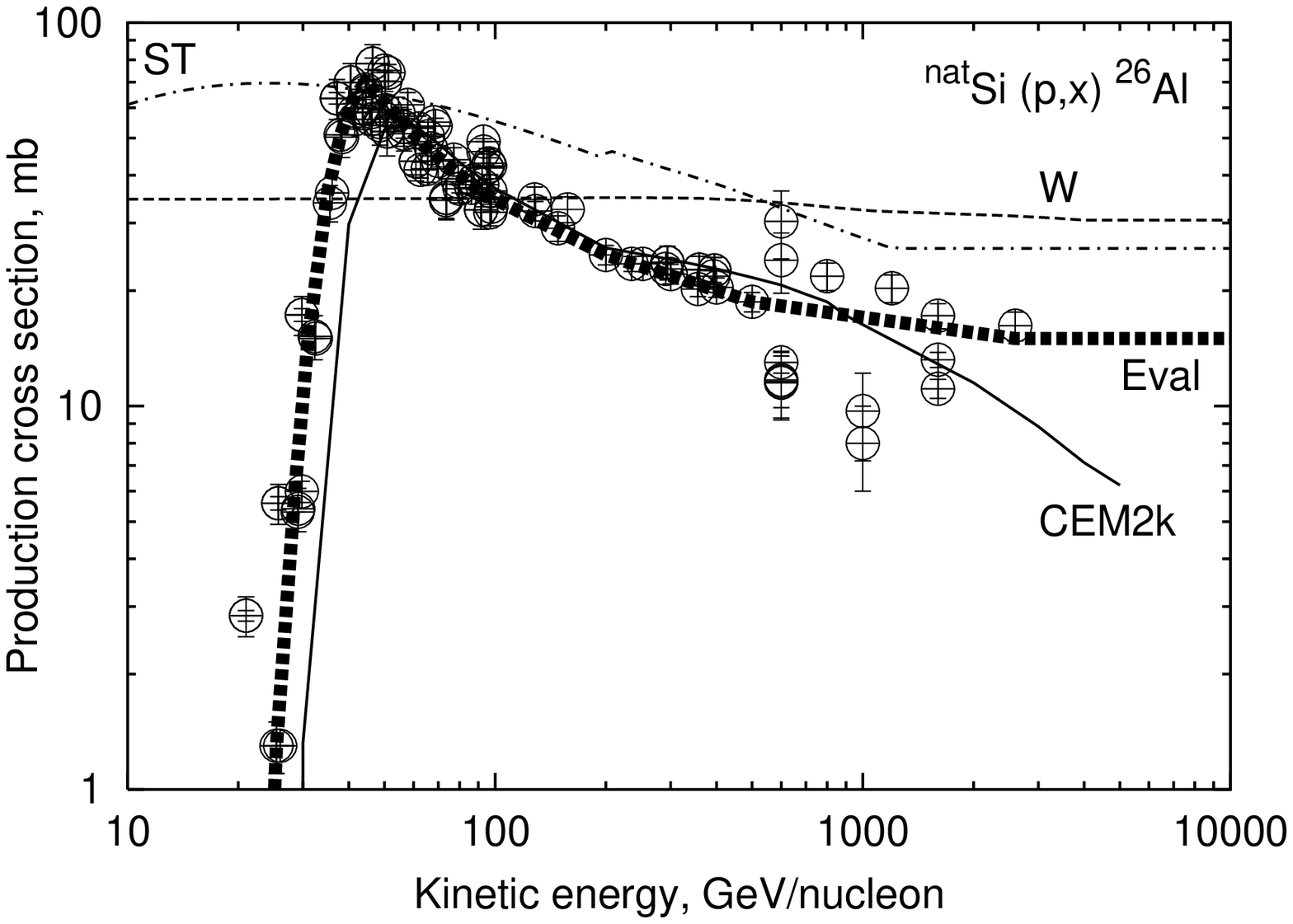} 
\vspace{-3\baselineskip}
\caption[fig1a.ps]{
Evaluated excitation function for the reaction $^{nat}$Si(p,x)$^{26}$Al
(thick dashed line) compared with experimental data from LANL T-16 
compilation (Mashnik et al., 1999) and results by CEM2k (thin
solid line) and phenomenological approximations by Webber et al. (1990)
(dashed line) and by Silberberg et al. (1998) (dot-dashed line).
}
\end{minipage}
\hspace{\fill}
\begin{minipage}[t]{88mm}
\centerline{\includegraphics[width=88mm]{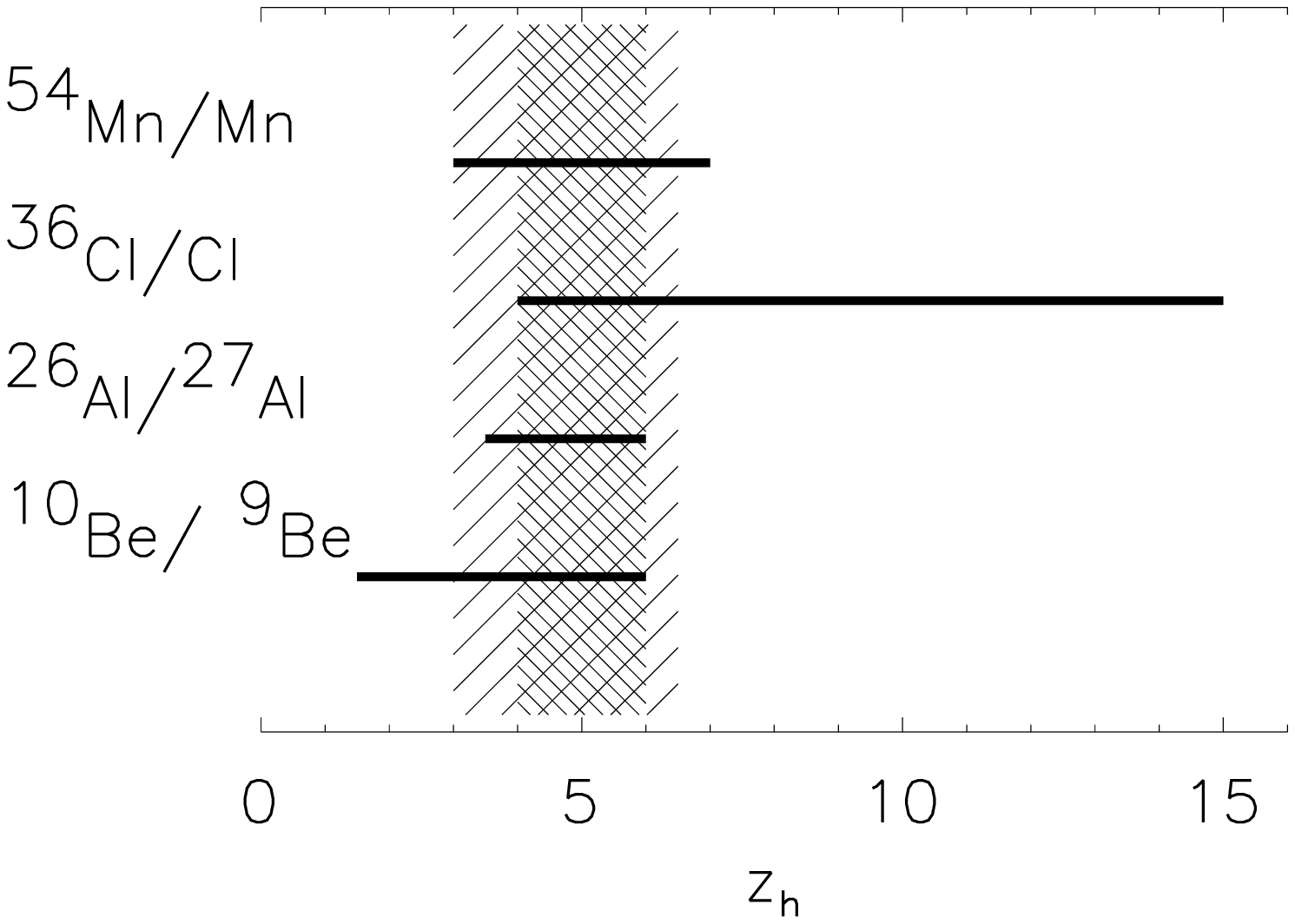}} 
\vspace{-10mm}
\caption[fig6.ps]{Halo size limits as derived in
Moskalenko et al. (2001)
from the abundances of the four 
radioactive isotopes and ACE data.
The ranges reflect errors in ratio measurements and source abundances. 
The dark shaded area indicates the range consistent with all ratios;
for comparison the range derived by Strong and Moskalenko (2001)
is shown by light shading.
}
\end{minipage}
\end{figure*}

\section*{ACKNOWLEDGEMENTS}
We thank 
Prof.\ Nakamura, Drs.\ Iwata, and Iwase for sending us
numerical values of their measured neutron
spectra and results of calculations with QMD and HIC.
This study was supported by the U.\ S.\ Department of Energy and by the
Moldovan-U.\ S.\ Bilateral Grants Program, CRDF Project MP2-3025.
S.G.M.\ and I.V.M.\ acknowledge partial support from a NASA
Astrophysics Theory Program grant.

\vspace{10mm}
{\noindent
E-mail address of S.G. Mashnik \hspace{2mm} mashnik@lanl.gov}\\
\\
Manuscript received 19 October 2002; revised \hspace{30mm} ; accepted
\end{document}